\documentclass[namedreferences]{solarphysics}
\usepackage[optionalrh]{spr-sola-addons} 
\usepackage{graphicx}        
\usepackage{color}           
\usepackage{url}             

\newcommand{\adv}{    {\it Adv. Space Res.}} 
 
\newcommand{\aap}{    {\it Astron. Astrophys.}}
\newcommand{\aaps}{   {\it Astron. Astrophys. Suppl.}}

\newcommand{\apj}{    {\it Astrophys. J.}}

\newcommand{\mnras}{  {\it Mon. Not. Roy. Astron. Soc.}}

\newcommand{\solphys}{{\it Solar Phys.}}

\begin{document}

\begin{article}

\begin{opening}

\title{The solar diameter series of the CCD Solar Astrolabe of the Observat\'orio Nacional in Rio de Janeiro measured during cycle 23}

\author{C.~\surname{Sigismondi}$^{a,b}$\sep
        S.C.~\surname{Boscardin}$^{a}$\sep
        A.H.~\surname{Andrei}$^{a}$\sep
       J.L.~\surname{Penna}$^{a}$\sep
       E.~\surname{Reis Neto}$^{a}$\sep
       V.A.~\surname{D'\'Avila}$^{a}$      
      }
\runningauthor{Sigismondi et al.}
\runningtitle{Rio de Janeiro solar astrolabe results}

   \institute{$^{a}$ Observat\'orio Nacional, Rio de Janeiro, Brazil
              $^{b}$ ICRANet, UPRA and Sapienza University of Rome, Italy email: \url{sigismondi@icra.it} \\
             }

\begin{abstract}

  The interest on the solar diameter variations has been primary since the scientific revolution for different reasons: 
  first the elliptical orbits found by Kepler in 1609 was confirmed in the case of the Earth, 
  and after the intrinsic solar variability was inspected to explain the climate changes.
  The CCD Solar Astrolabe of the Observat\'orio Nacional in Rio de Janeiro made daily
measurements of the solar semi-diameter from 1998 to 2009, covering most of the cycle 23, 
and they are here presented with the aim to evidence the observed variations.
Some instrumental effects parametrizations have been used to eliminate the biases appeared in morning/afternoon data reduction.
The coherence of the measurements and the influence of atmospheric effects are presented, to discuss the reliability 
of the observed variations of the solar diameter.
Their amplitude is compatible with other ground-based and satellite data recently published.

\end{abstract}
\keywords{Solar Diameter; Instrumentation and Data management; Instrumental effects;Atmospheric effects.}
\end{opening}

\section{Solar diameter historical measures from 1600 to 1900}

The interest on the solar diameter variations has been prioritary during the last four centuries
of astronomy with and without the telescope.
During 1600 the verification of the Keplerian hypothesis of elliptical Earth's orbit was possible 
thanks to measurements of the seasonal variations of the angular diameter of the Sun made by 
G. D. Cassini and G. B. Riccioli from 1655 to 1666 at the {\sl Heliometer} of S. Petronio in Bologna.
This was a pinhole solar telescope\footnote{The use of the word {\sl telescope} for an optics-less instrument like a 
giant pinhole camera realized into a church provided by a meridian line accurately draft on the ground,
may appear awkward. But the Greek word's meaning is related to the capability of seeing at far objects, 
and the implicit possibility to have magnified images, and this is possible with pinhole instruments.} 
of the Basilica of St. Petronius in Bologna \cite{Manfredi}.
Later in 1700 the interest of the astronomers was more focused on the Earth's axis orientation,
with the variation of the obliquity $\epsilon$. To measure this parameter it was necessary to know better the position of the 
solar center, than the diameter. The giant pinhole telescopes \cite{Heilbron} could operate on a complete yearly range of zenithal distances of $\sim 47^{o}=2\times\epsilon$ without 
optical distortions, excepted the atmospheric refraction, and for this reason these instruments continued successfully their operations
until the end of the century  even if their imaging capability was comparatively lower than usual  telescopes.
The heliometer as we know nowadays was invented in 1743 by S. Savery \cite{Short1753} and built by J. Dollond ten years later, but
the solar diameter has been measured with timing meridian transits at the Greenwich Observatory, starting with N. Maskelyne from 1765 to 1810,
and the influence of a variable {\sl personal equation} with the age, either in reaction time either in contrast sensitivity, 
has been recognized as the most reliable cause of the recorded variations.
 
The heliometric method, which does not deal with reaction times, but only with the personal contrast sensitivity, was considered more accurate than \cite{Auwers1890}. 
The nineteenth century was the golden age of the solar astrometry, and the heliometers built by J. Fraunhofer 
were so accurate that were used by F. W. Bessel for the measurement
of the first stellar parallax \cite{Bessel1838}.
The measurements of the solar diameter variations
became a routine with specific instrument designed for, and the observatories involved in the meridian transit measurements 
were also Neuchatel, Oxford, Washington \cite{Auwers1890}, and Rome-Campidoglio \cite{Gething1955}.
The current standard value of the solar semi-diameter has been fixed to 959.63 arcsec\footnote{This angular dimension corresponds to 696000 Km at 1 AU.} by A. Auwers in 1891 \cite{Auwers1891}.
The words of A. Secchi in the book "Le Soleil" expressed clearly the status of art at the fall of the century \cite{Secchi1875}:
{\sl What strikes even more is to see that, despite
the variety of methods and instrumental perfection,
the measurement of the solar diameter made very little progress.}\footnote{Ce qui frappe encore davantage, c'est de voir que, malgr\'e
la vari\'et\'e des m\'ethodes et la perfection des instruments,
la mesure du diam\`etre solaire a fait bien peu de progr\`es (Secchi, 1875 p. 210).}

\section{Solar diameter and solar figure in the last century}

The 1900 opened with a further development in the heliometric technique, made in Goettingen \cite{Schur05}.
The splitted lens was substitued by a prism in front of the single lens. The heliometric angle was therefore fixed 
by the aperture angle of the prism, and the distance between the two images was the only variable to be measured, being fixed the focal length.
This configuration was used again in the Solar Disk Sextant (SDS) experiments in the 1990s decade \cite{Egidi06} and in 2009 and 2011 \cite{Sofia2013}. 
SDS flew seven times on the top of the stratosphere where no turbulence and refraction effects act. 
The disadvantage is that the measurements can last only up to 9 hours each flight, 
but it allows comparison of diameter measures across decades. 
The variations of the diameter observed with this experiment of metrological quality are of 0.2 arcsec, 
while the typical estimated uncertainty of each measure is 0.02 arcsec.
The heliometric angle is stable and it has been verified within 0.1 arcsec: this is the better accuracy of the measure of the wedge angle
made by multiple internal reflections of a LASER beam through the two faces of the objective prism. 

The figure of the Sun also become interesting for the physicists, starting from the studies
on General Relativity in order to understand the contribution (of classical Newtonian physics) of the solar oblateness 
to the perihelion advancement of Mercury \cite{Sigismondi11,Sigismondi14}.
A new interest in the solar diameter variation sprung in 1978 when J. Eddy evidenced secular variations 
of the solar diameter on the basis of the annular eclipse observed by Clavius in 1567. 
This eclipse with the standard solar radius should have been total, while the solar photosphere exceeded the lunar limb to show the observed ring.

In the same years (1974-75) a long lasting observative campaign of monitoring solar diameter started in Calern, France\cite{Laclare83} 
and in Brasil (S$\tilde{a}$o Paulo \cite{Emilio05} and Rio de Janeiro \cite{Penna96}) by using the astrolabes of Danjon \cite{Danjon55}
opportunely modified to observe the Sun with silicon density filters with dielectric multilayer coating to reduce the luminosity of the Sun \cite{Laclare83}.

Historical data from meridian transit were analyzed \cite{Wittmann77} and compared with modern measurements \cite{Bianda00} made in Locarno and Iza$\tilde{n}$a/Tenerife.
The last generation of the solar astrolabes, with CCD and varying prisms and several automated controls
 is represented by the instruments of R2S3 network, described in the following paragraph.

\section{The solar astrolabes network}

Several groups worked on the measurements and
observed variations of the solar semi-diameter in the last four decades. 
A network of them, using metrological astrolabes
adapted to CCD solar observations and equipped with a varying prism and a rotating shutter,
the R2S3, a French acronym standing for "Sun radius ground survey network"
was established in the last decade \cite{Delmas06}.
The Danjon astrolabe in use at Rio de Janeiro Observatory, 
has also been equipped with a CCD and a varying prism 
in order to perform systematic measurements of the solar diameter's variations.
The software used in the image treatment, and for the identification of the 
solar limb was developed for the French astrolabes in 1998 \cite{Sinceac98}.
The same software was used also for the DORAYSOL astrolabe at Calern/OCA,
operating from 1999 to 2006 \cite{Morand10}. 

The timing accuracy in the images acquisition is also metrological, better than 15$\mu$s
(corresponding to $\sim 0.2$ milli arcsec in the solar radius)
being guaranteed either in Calern and in Rio de Janeiro by the reference time service 
spread by these Observatories.

Other Observatories participating to this network are san Fernando in Spain, Antalya in Turkey\cite{Kilic2005} and Tamanrasset in Algeria. 
S$\tilde{a}$o Paulo adn Santiago de Chile astrolabes did not work with variable prisms angles.

\section{The astrolabe in Rio de Janeiro}

At the Observat\'orio Nacional/MCT (ON) in Rio de
Janeiro (Lat=-22deg53'42'', Long.= +2h52m53s.5, h=33m) the series
of solar semi-diameter measurements started in 1997, in
some measure provoked by systematic biases found during
several preceding years of astrometric observations of the
Sun. For the ensuing semi-diameter long term campaign the
instrument underwent several modifications. The most important
were the installation of a variable angle front prism
enabling the continuous observations between the zenith distance
of $25^o$ and $56^o$, the concurrent installation of a moving
density filter, and the installation of a CCD camera, which
allowed the observations to become fully freed of personal
equations. A complete account of this setup, in the major
measure developed and adopted by the Calern solar group,
was presented in several papers \cite{Sinceac98}; \cite{Jilinski1998}; \cite{Jilinski1999}; \cite{Penna2002}. 

Here we analyze the period from 1998 up to 2009. The first year of observations, 1997,
was discarded as discussed in \cite{Reis-Neto2003}. 
The instrument went through a major upgrading in 2004 \cite{Boscardin2011}.

The data file here concerned comprises the rise, crest, subduing of the solar cycle 23
and the beginning of the long minimum between solar cycles 23 and 24, which resulted the longest
since the Dalton minimum of 1810 \cite{Usoskin2008}.

\subsection{The solar filter}

As all solar astrolabes, there is a solar filter blocking the majority of the solar radiation.
This is a neutral density filter with a metal coating, on a double glass, which thickness is 3.5 cm. The transmittance of this filter is $10^{-4}$.
This filter is mounted in a fixed position, and oriented toward the Sun with a tolerance of a few degrees,
to avoid internal reflections and ghost imaging.
In front of CCDs there is a couple of filters for selecting the passing waveband of light from 523 nm to 691 nm. 
The maximum transmission of this pair of filters is at 563 nm, and this is the effective wavelength of the observation.
For reference the wavelengths selected to observe the
solar photospheric continuum by the Picard satellite mission SODISM experiment are 535.7, 607, and 782.0 nm and bandwidths respectively of 0.5, 0.7 and 1.6 nm; 
moreover there are other windows within the Ca II Fraunhofer lines at 393 nm, 
and over a broad wavelength range of 7 nm centred
at 215 nm, which includes many Fraunhofer lines, to study their effects on the solar limb
shape and the diameter as a function of solar activity \cite{Meftah2014}. 

The models of the solar limb luminosity profiles are studied in deep detail for understanding the response of two instruments: 
the Solar Disk Sextant is centered at 610 nm with 100 nm of bandwidth \cite{Sofia2013} 
and the Heliometer of the Pic du Midi at $782 \pm 1.6$ nm \cite{Thuillier2011}.

The filters used in the solar astrolabe of Rio overlap the same wavelengths of the Solar Disk Sextant, 
and if there is any influence of some emission lines on the measured diameter, it is the same for both instruments.

\subsection{Solar astrolabe principles and the azimut variation during an observation}

The method the solar astrolabe observations minimizes the influence of the optical distortions.

The principle of the measurements uses two images of the
Sun: one is said direct while the other follows a path that
reflects on a horizontal basin of mercury. To each image, an arc of
parabola is adjusted to fit the steepest descent of the radial luminosity profile (inflexion point) 
to define the solar limb \cite{Golbasi2001}. 
In this way the influence on personal equation due to the contrast perception, is eliminated.
The contact between direct and reflected image is always observed at the center of the field of view, 
in such a way the optical path of the rays through the telescope is the same, and all distortions behave like systematical errors.

\begin{figure}    
\centerline{\includegraphics[width=0.5\textwidth,clip=]{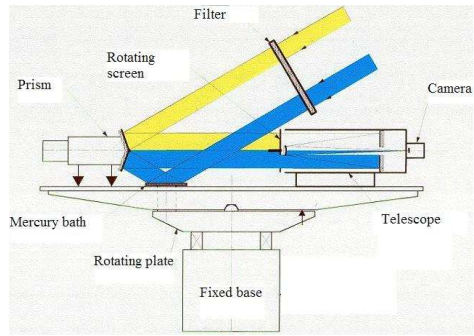}}
\caption{Number of observations of the solar semi-diameter from 2001 to 2009, with the Rio de Janeiro astrolabe.}
\end{figure}

Since the measurement is not instantaneous, as in the case of heliometric images of the Solar Disk Sextant or the Reflecting Heliometer of Rio de Janeiro, 
in the two instants of the contacts of the solar limbs with the almucantarat (circle of a given altitude in the sky) 
the light rays pass through different regions of the atmosphere, which is turbulent at all length scales.
When the refraction is slightly different either on the same altitude circle or in the same direction and different times 
we are in presence of the {\it anomalous refraction} \cite{Taylor13}, 
and this has been already evidenced for solar observations by \cite{Sigismondi2012}. 

Moreover the second contact with the same almucantarat occurs at a slightly different azimut, 
yielding again a modification of the optical path in the atmosphere.

\begin{figure}    
\centerline{\includegraphics[width=0.5\textwidth,clip=]{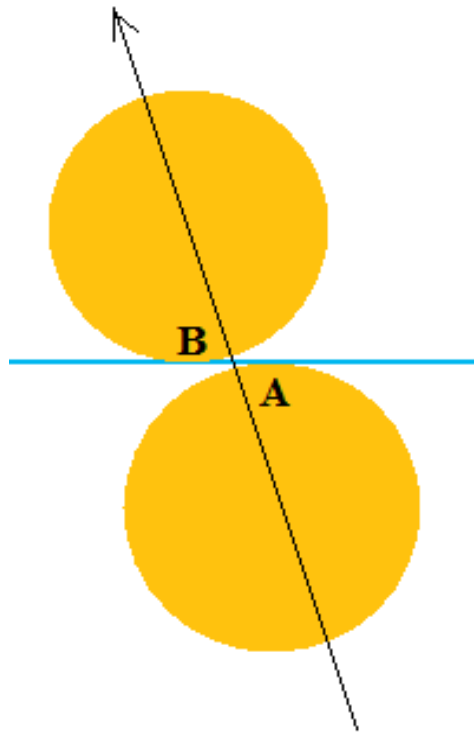}}
\caption{Number of observations of the solar semi-diameter from 2001 to 2009, with the Rio de Janeiro astrolabe.}
\end{figure}

A random variation of nul average is expected to occur in this case.
This fluctuation is tacitly considered negligible and expected to be average to zero over several measurements. 

Finally the heliolatitudes observed in Rio de Janeiro, thanks to its tropical location, cover the whole solar figure in a semi-annual cycle;
this is not possible for Observatories located in other latitudes: for them the measurable heliolatitudes are limited.

\subsection{Data qualification of the Rio Astrolabe}

The solar semi-diameter series was observed from August 1st, 1998 to November 30th, 2009
(Figure 1). It comprises more than 19000 observations, on all heliolatitudes,
with mean internal error of 0.20 arcsec and standard deviation
of 0.57 arcsec. 

\begin{figure}    
\centerline{\includegraphics[width=0.5\textwidth,clip=]{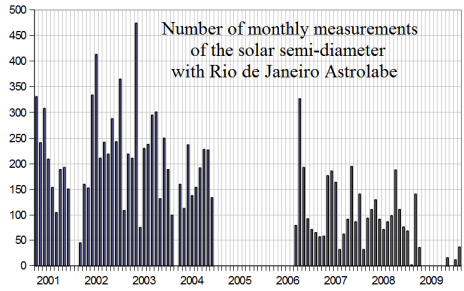}}
\caption{Number of observations of the solar semi-diameter from 2001 to 2009, with the Rio de Janeiro astrolabe.}
\end{figure}

The observations have been made daily, to an average
of 20 observations per duty day, well
distributed throughout the whole year. The gaps verified in
the series between September 21st and December 19th, 2001
and between June 1st 2005 and August 10th 2006 
were due to maintenance of the apparatus. The observations
are taken on sessions before and after the meridian transit. 

The relaxation time of the spring used to regulate the prism angle was different from the morning 
to the afternoon, due to different temperatures, and this created a bias between these data.
A linear model of the relaxation process of the spring acting on the variable prism, 
fully accounted for this effect, eliminating the bias.

\begin{figure}    
\centerline{\includegraphics[width=0.5\textwidth,clip=]{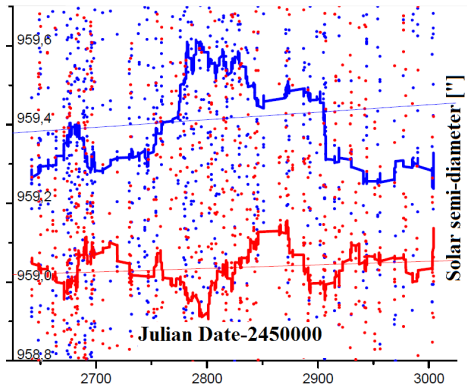}}
\caption{Asymmetry in East and West measurements before the corrections (see text).}
\end{figure}

Therefore there is no significant difference between anti e post meridian measurements, 
unlike for DORAYSOL \cite{Morand2010}
which is the instrument more similar to the Rio Astrolabe. 

The raw data were corrected from effects related
to the observation conditions: the air temperature, its
first derivative, the Fried factor and the standard deviation
of the points of the adjusted parabole to the directly observed solar edge
\cite{Boscardin2005} and \cite{Boscardin2011}. 
The Fried factor was obtained from the observation data cf. \cite{Lakhal1999}.

\begin{figure}    
\centerline{\includegraphics[width=0.5\textwidth,clip=]{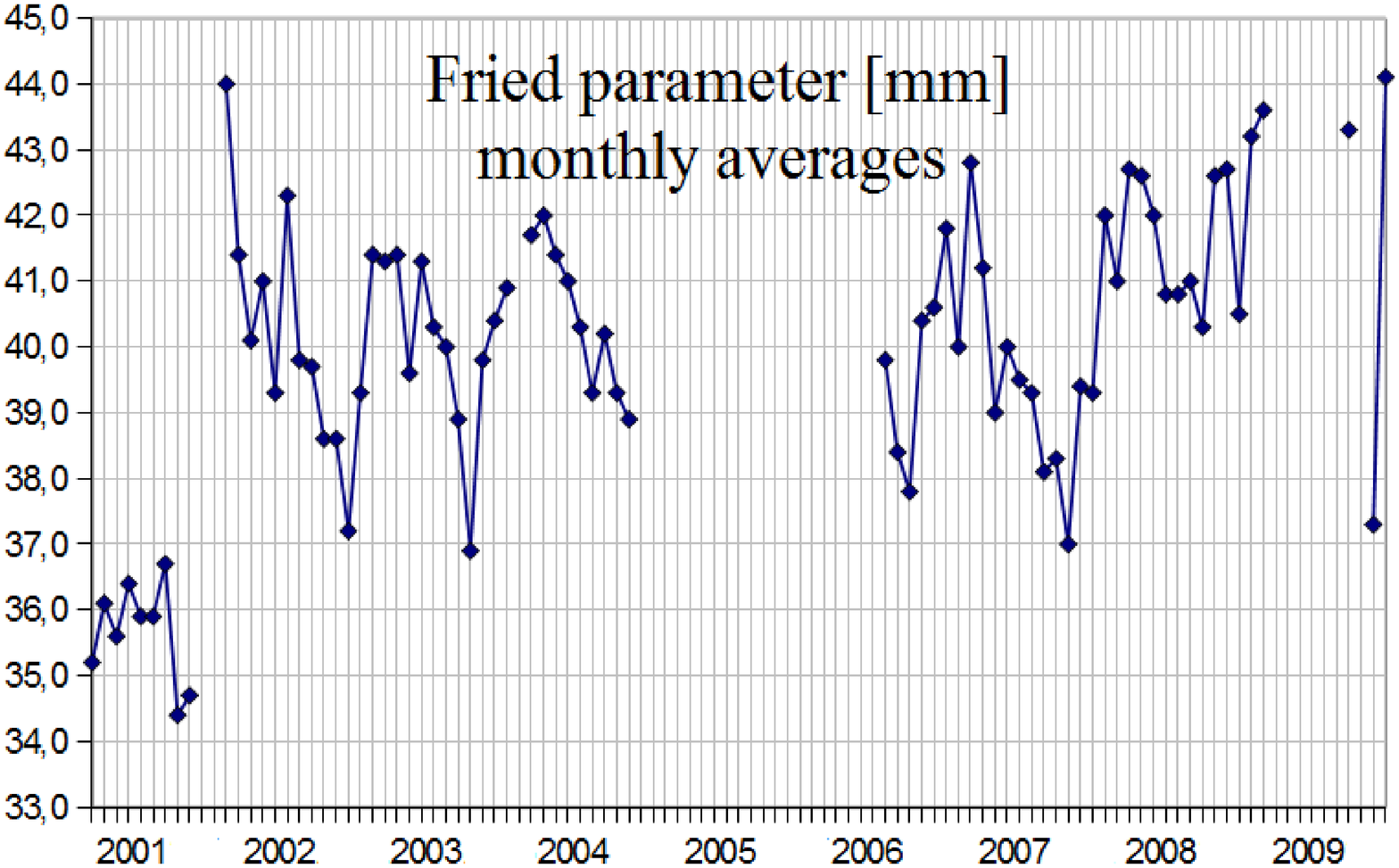}}
\caption{The Fried factor as measured with the astrolabe in Rio de Janeiro: monthly averages.}
\end{figure}

Further instrumental conditions were inspected in order
to detect effects caused by any instability of the objective
prism \cite{Reis-Neto2002} and from the lack of leveling of
the astrolabe that could cause errors as function of the observed
azimuth \cite{Boscardin2005}. 
The standard deviation
of the data before and after introducing appropriate correcting parameters remained unchanged, 
showing that all corrections
applied were negligible and did not introduce any spurious long-
term modulation upon the series. 
Seasonal or annual effects
on the raw measurements are very small, coherent with the
standard refraction theory. 

\section{Correlations diameter-activity}

The final series of solar semi-diameter
values was correlated against the series of solar
activity parameters in the common period.

\begin{figure}    
\centerline{\includegraphics[width=0.5\textwidth,clip=]{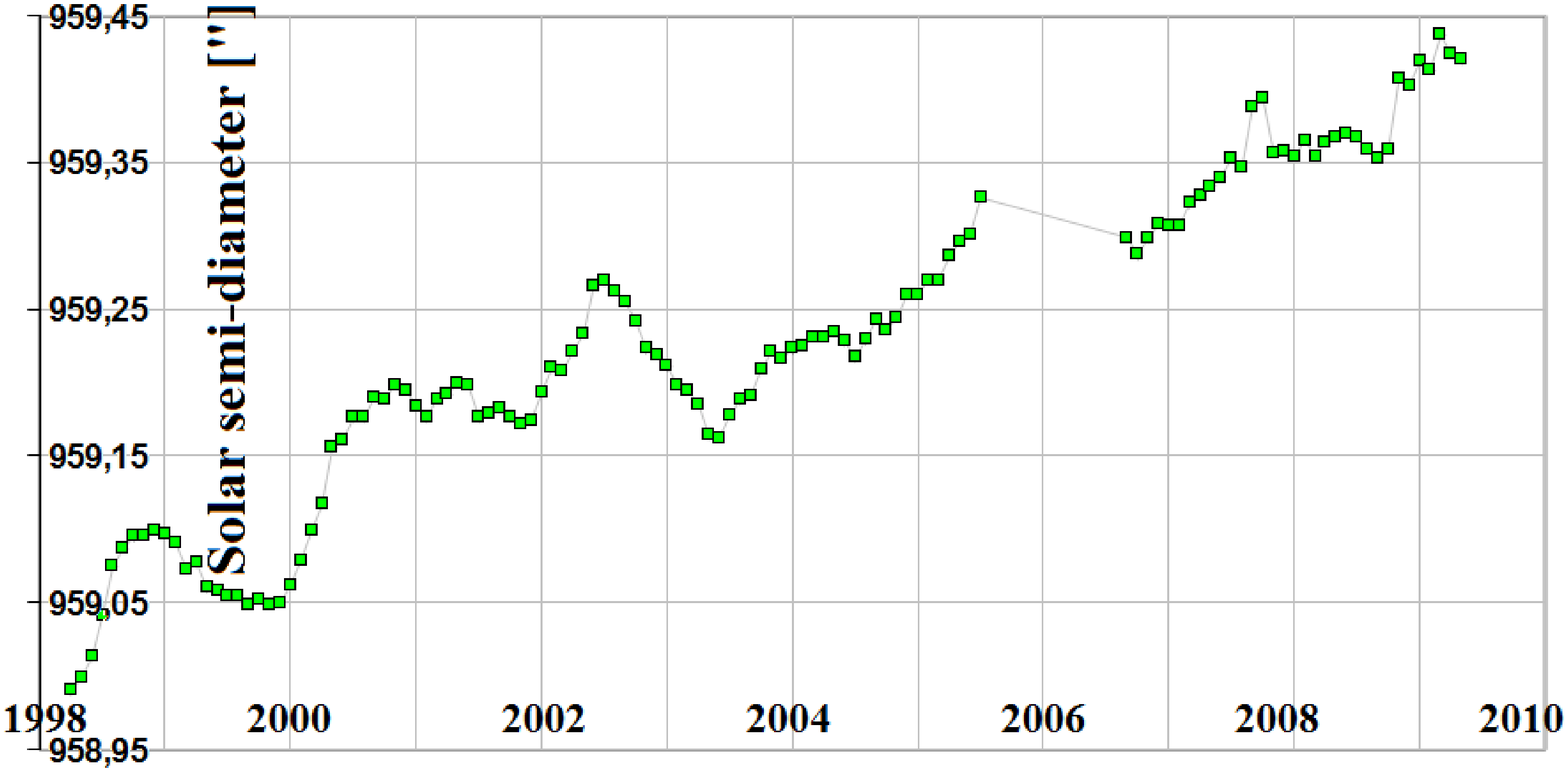}}
\caption{The series of the solar semi-diameter measured with the astrolabe in Rio de Janeiro: monthly averages from 1998 to 2009.}
\end{figure}

The observations along a whole solar cycle made at the Rio Astrolabe 
thanks to their large quantity and even spreading, offered the conditions to statistically compare the
observed solar semi-diameter variations against estimators
of the solar activity, which could also be expressed as continuous
daily series. The estimators used were the sunspot
count number and its proxy, the 10.7cm radio flux, both
sensing the photosphere state; the total solar irradiance and
the strength of the integrated solar magnetic field, sensing
directly the solar cycle age, and finally the flare index
to assess the major solar outbursts. 

All the estimator data
were retrieved from the National Geophysical Data Center -
NGDC. The hypothesis that the variation of the solar semidiameter
could be linked to the solar activity was checked
calculating the correlations between the solar semi-diameter
series against each one of those estimators. Afterwards, the
correlations were recalculated, taking pairs of correlated series,
and allowing parametric time delays between them. This
may point to interconnected phenomena, either with some
time delay between them, or even a causal relationship. In
order to get a broader picture, the correlations among the
estimators were likewise calculated \cite{Boscardin2011}.

\begin{figure}    
\centerline{\includegraphics[width=0.5\textwidth,clip=]{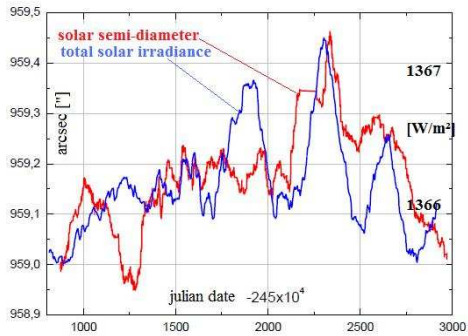}}
\caption{The solar semi-diameter series is correlated with the solar activity (NGDC data 1998-2003).}
\end{figure}

\subsection{Methodology of the statistical analyses}

The large number of points and regular distribution of the
measurements of the solar parameters, both regarding the
semi-diameter and the estimators of the activity, prompted
to the use of the linear correlation coefficient as the measure
of relationship. It is robust in this case because with it there
is no implicit assumption of variation of the semi-diameter,
nor on how the activity estimators varied. In the following
subsections, the correlation coefficient is calculated for increasing
binning periods. They divide the full span from 12
up to 144 equal sampling intervals. That is, roughly corresponding
to binning periods from one month to one year. In
this way, the correlations cover a range from the more detailed
to the broader features. The correlation coefficient is
calculated independently for each binning (aliasing) period. But, in order
not to privilege particular intervals, the beginning dates
of the sampling have been displaced backward and forward within 10 days,
with the final result behaving like the correlation between 10 days running averages of the considered measurements.

\section{Discussion: solar astrometry with astrolabes and solar structure's theory}

All long-term series of R2S3 astrolabes and SDS measurements have been put 
in correlation with the cycles of solar activity, and among different instruments 
these correlations disagree concerning phase and amplitude. 

Correlations between the observed series of the solar diameter 
and other parameters concerning the solar activity: namely spots and radio flux at 10.7 cm, have been presented in Boscardin (2011). 
They can be analyzed and discussed in the light of the last theoretical works published
on the solar structure.

Some observational and theoretical reasons 
explain the departures form sphericity of the solar figure\cite{Badache2006}. 
Although there is not a comprehensive cause to model 
the observed variations of the photosphere
diameter up to now, some hypothesis have been put forward.
\cite{Spruit2000} and \cite{Dziembowski2002} showed how the solar irradiance can be influenced by the magnetic field driving the solar cycle:
higher irradiance is explained by a corrugated surface rendering the Sun a more effective radiator.
This study predicts only micro-variations ($\le 10^{-6}$) for the solar diameter. The observed ones are up to hundred times larger. 
\cite{Sofia2005}; \cite{Andrei2006};\cite{Badache2007} took advantage of these observational
evidences to improve the already extremely detailed solar theory. 

Ultimately, thus, although it is not in the scope of this article,
an approach towards a unified description of the observations
should pave the way to place the solar semi-diameter
variations within the general frame of the solar photosphere
description.

Are these the reasons to present here the semi-diameter series of the Rio astrolabe, without any further correlation study, since
other long-term series have been recently published, and the experimental panorama is worth to be completed.
We remember that the possibility to monitor continuosly over decades the solar diameter, given by the ground-based experiments like the solar astrolabes  
and recently the Reflecting Heliometer of Rio de Janeiro \cite{Andrei2013,Sigismondi2013,Sigismondi2013b}, is the only way to extend backwards of some decades, 
with the present study, and forward beyond the lifetime of any satellite,
the knowledge about the variability of the Sun.
The importance of such ground-based experiments and their results remain unsubstituable.

\subsection{Statistical limits or real diameter fluctuations?}

The contrast between helioseismology restrictions on diameter's variations, based on the shifts of the centroids frequencies in the spectrum of solar 
oscillations through the solar cycle (e.g. a bigger sphere oscillates at a lower frequency) manifesting spherically symmetric changes in the Sun, 
and the observational results on the solar diameter 
has being evidenced in the last decades with several methods:

\begin{itemize}

\item {the solar astrolabes which marked the standard for subarcsecond solar astrometry after 1975;}

\item{total and annular solar eclipses of the past and present decade \cite{Sigismondi2009,Raponi2012} 
and the last transits of Mercury (2003 and 2006) and Venus (2004 and 2012): a significant refinement 
in measuring the solar diameter using solar lunar and planetary ephemerides with more accurate timings of 
these celestial alignments and new analyses are moving forward this field of astrometry down to 0.01 arcsec of accuracy;}

\item{satellite measurements, like Picard and SOHO, with the problem of their limited lifetime.}

\end{itemize}

The annual averages of the series of hundred years of meridian transits in Greenwich and Campidoglio Observatories \cite{Gething1955} 
showed scatters from one year to the following
often larger than the standard deviation of the yearly average.
The same behaviour visible in the astrolabe data, binned either over one year or one month, suggested to us to further investigate
on the source of these deviations.

The Reflecting Heliometer\cite{Andrei2013,Sigismondi2013,Sigismondi2013b} continued the measurements on the local seeing effects, getting the first spectra up to 3 mHz cite{Sigismondi2013c}.
The IRSOL telescope recorded firstly seeing spectrum up to 0.3 mHz: in both cases the power spectrum showed energy 
at $2\div5$ minutes level corresponding to the durations of either meridian or almucantarat transits.
 
The real origin of these scatters could be an underestimate of the systematic errors of the measurements.
But since the astrolabe reduced these errors to the minimum theoretically possible, 
unless invoking unprobable atmospheric phenomena acting only on the line of sight of the telescope,
we have only to define such effects as non-Gaussian.
The energy of daytime atmospheric turbulence at $2\div5$ minutes is as big as 1 arcsec of amplitude, producing such observed fluctuations.
If the atmospheric turbulence does not explain these fluctuations they are real oscillations of the solar diameter.

The use of a running average smoothes always all gaps, and this happen in the data presented in the monthly averages presented in the figure. 
After astrolabes and SDS we know that the amplitude of solar semi-diameter variations is 0.1 arcsec.
The results of the measurements of 1850-1950 meridian transits showing an amplitude of 0.5 arcsec seem to be ruled out. 
But we have to bear in mind that the Sun, meanwhile, undergo significant modulations of its activity, having a grand maximum lasted from 1960 to 2000 
(the {\it Eddy maximum}) and a rather pronounced minimum after 2009 lasted 700 days 
followed by a 24th cycle lower than the previous ones, considered as a preludium of a new grand minimum \cite{Penn2006}.

On the other hand the Sun at grand minimum of activity cannot be smaller than any normal minimum \cite{Dziembowski2002}
since there is no departure from sphericity at each minimum of activity, 
and with the rising of the activity the aspherical components of the oscillations increase.  
In this view the Sun at maximum activity has the maximum asphericity, rendering the surface corrugated and a more efficiently radiator.
Negligible variations of the solar radius occur in this case from minimum (even grand minimum) to the maximum of solar activity.

\section{Conclusions}

The aforementioned words of A. Secchi in 1875 could be applied also nowadays, even if meanwhile the errorbars 
of these measurements have been reduced of more than a factor of 10-100.
The influence of atmospheric turbulence and the environmental parameters can affect a single measurements with perturbations 
bigger than the real variations of the solar diameter.
The era of satellites opened the way to the milliarcsecond solar astrometry, but for the knowledge of the solar diameter fluctuations
on timescales longer than the solar cycle, their lifetime is very short and their operational cost bigger than astrolabes and heliometers.

The Solar Heliospheric Observatory SOHO has the longer operating time until now, but its instruments were not designed for astrometry.
{\it Nonetheless its data have been exploited to show that the solar diameter has not changed within a few milliarcsecond per year...}
(Delmas et al. (2006) p. 1567 p. 4.)

Because of the nature of planetary transit measurements of solar diameter, an excellent angular resolution is achievable by a good timing resolution.
Hence the Mercury transits of 2003 and 2006 observed by SOHO provided the occasion 
to measure the solar diameter based only on timing accuracy  independently on angular references \cite{Emilio12}.
Therefore these measurements made with SOHO in 2003 and 2006, 960.08 arcsec, can be fairly compared with the analogous 
measurements made by PICARD satellite during
the transit of Venus of 2012 in the same waveband, 959.86 arcsec, 
concluding that the variations of the diameter from 2003 to 2012 has been $0.22\pm0.10$ arcsec.
This is in perfect agreement with the variations measured by the Rio Astrolabe during the cycle 23, and this may explain directly 
the scatter between the yearly averages found in the aforementioned series of Greenwhich, Campidoglio, Rio de Janeiro and Calern.

The astrolabe of Rio de Janeiro allowed to monitor the solar during more than a decade without the costs 
of a space mission and its limited operating lifetime.
Future results from Hinode, SDO and PICARD satellite can better elighten this field of research, 
and instruments like PICARD-Sol and the Reflecting Heliometer of Rio de Janeiro
will help to maintain the 0.01 arcsec of accuracy in solar astrometry typical of the satellites 
to ground-based observations, to prolong the series on the solar diameter for the years to come.

The measurements of the solar diameter made with the solar Astrolabes have been debated since many decades. In this paper all original data of the observations made in Rio de Janeiro over the whole cycle 23 have been here presented as appendix from \cite{Boscardin2011} and discussed.
The scope was also to share with the international heliophysics community these data with the problems of their interpretations. Since this kind of problems are common with all ground-based observations \cite{Gething1955} their final solution will take full advantage also of these data, to verify, for example, the hypotheses on the possibility that the solar activity could influence indirectly the measurements as some scholars claimed \cite{Rozelot2003}.

\begin{acks}
C.S. thanks J.P. Rozelot and W. Dziembowski for fruitful discussions and suggestions.
\end{acks}

\end{article} 

\end{document}